\documentclass[aps,twocolumn,superscriptaddress, showpacs,prl]{revtex4}
\usepackage{amsmath,latexsym,amssymb}
\usepackage{amsmath}
\usepackage{latexsym}
\usepackage{amssymb}
\begin{document}
\title{Rotational invariance as an additional constraint on local realism}
\author{Koji Nagata}
\affiliation{National Institute of Information and Communications 
Technology, 4-2-1 Nukuikita, Koganei, Tokyo 184-8795, Japan}
\author{Wies{\l}aw Laskowski}
\affiliation{Instytut Fizyki Teoretycznej i Astrofizyki
Uniwersytet Gda\'nski, PL-80-952 Gda\'nsk, Poland}
\author{Marcin Wie{\'s}niak}
\affiliation{Instytut Fizyki Teoretycznej i Astrofizyki
Uniwersytet Gda\'nski, PL-80-952 Gda\'nsk, Poland}
\author{Marek {\. Z}ukowski}
\affiliation{Instytut Fizyki Teoretycznej i Astrofizyki
Uniwersytet Gda\'nski, PL-80-952 Gda\'nsk, Poland}
\affiliation{Institute f\"ur Experimentalphysik, Universit\"at Wien, A-1090, Austria}
\begin{abstract}
Rotational invariance of physical laws is a generally accepted principle. 
We show that it leads to an additional external constraint on local realistic models
of physical phenomena involving measurements of multiparticle spin $\frac{1}{2}$ correlations. 
This new constraint rules out such models even in some situations in which 
standard Bell inequalities allow for explicit construction of such models.
The whole analysis is performed without any additional assumptions on the form 
of local realistic models.
\end{abstract}

\maketitle

Local realism is a foundation of classical physics \cite{bib:Einstein, bib:Bell, bib:Redhead, bib:Peres3}. It is a conjunction of {\em realism}, i.e.
assumption that physical systems posses properties, irrespective whether these are
measured or not, and {\em locality}, which is the assumption of a finite
speed of influences (i.e. it is a consequence of special relativity).
Quantum mechanics does not allow a local realistic interpretation. 
The quantum predictions violate Bell inequalities \cite{bib:Bell}, 
which are conditions that all local realistic theories must satisfy.
Many of the recent advances in quantum information theory suggest
that the highly-non-classical properties of quantum states that lead to violations 
of Bell inequalities can be used as a resource to achieve success in some 
tasks, which are classically impossible. As examples can serve quantum cryptography and
quantum communication complexity \cite{EKERT, BZPZ}.
Therefore as the impossibility of existence of classical models for some 
processes leads to various quantum informational applications it is
important to learn what are the ultimate bounds for such models.

Here we aim to show that the fundamental property of the known laws of physics, 
their rotational invariance, can be used to find new bounds on local realistic description of phenomena involving multiparticle spin correlation \footnote{As it is well known, the rotational invariance of physical laws leads to angular momentum conservation.}. 
Our thesis is as follows. 
Assume that one has a correlation function for a given process involving spin measurements by several parties. 
This correlation function has a form which is rotationally invariant (see equation (\ref{et})).
One wants to build a local realistic model for the correlation function. It turns out 
that the demand that the resulting correlation function must be rotationally invariant
leads to a new type of Bell inequality, which restricts additionally possible local realistic models.
Further, even if ``standard" two setting Bell inequalities allow local realistic models for the given set of data (i.e., a set of correlation function values obtained in a Bell type experiment),
the new restriction, derived from rotational invariance, can invalidate such models, for some range of parameters.  

An important note here is that we do not impose rotational invariance on the
local realistic models themselves. This would be equivalent to making some additional assumptions about their form.
Instead we assume that the correlation functions observed in a Bell-type experiment have
the property of rotational invariance, which is something to be expected, due to the fundamental role of this principle. 
Note here, that of course, the quantum correlation function has an explicitly rotationally invariant form.

Assume that we have a set of $N$ spins $\frac{1}{2}$. Each of them is in a separate laboratory. As is well known the measurements (observables) for such spins are parameterized by a unit vector $\vec n_i$ (direction along which the spin component is  measured). The results of measurements are $\pm 1$ (in $\hbar/2$ unit). One can introduce the ``Bell'' correlation function, which is the average of the product of the local results:
\begin{equation}
E(\vec n_1, \vec n_2, ..., \vec n_N) = \langle r_1(\vec n_1) r_2(\vec n_2) ... r_N(\vec n_N) \rangle_{avg},
\end{equation}
where $r_j(\vec n_j)$ is the local result, $\pm 1$, which is obtained if the measurement direction is set at $\vec n_j$. Any theory which is satisfying the principle of rotational invariance must predict the following structure of $E(\vec n_1, \vec n_2, ..., \vec n_N)$ (provided one assumes that $E$ is linearly dependent on $n_j$'s, just like it is the case for quantum mechanical correlation functions \footnote{We choose this specific class of rotationally invariant formulas (\ref{et}) because our final aim is to compare
the possible local realistic models with the predictions of quantum mechanics. An analysis of possible more complicated forms of possible correlation functions would not contribute anything to this final aim, except for some unnecessary mathematical complication.}):
\begin{equation}
E(\vec n_1, \vec n_2, ..., \vec n_N) = \hat T \cdot (\vec n_1 \otimes \vec n_2 \otimes ... \otimes \vec n_N),
\label{et}
\end{equation}
where $\otimes$ denotes the tensor product, $\cdot$ the scalar product in $\mathrm{R}^{3N}$ and $\hat T$ is the correlation tensor given by
\begin{equation}
T_{i_1...i_N} \equiv E(\vec x_{1}^{(i_1)},\vec x_{2}^{(i_2)}, ..., \vec x_{N}^{(i_N)}),
\label{tensor}
\end{equation}
where $\vec x_{j}^{(i_j)}$ is a unit directional vector of the local coordinate system of the $j$th observer; $i_j = 1,2,3$ gives the full set of orthogonal vectors defining the local Cartesian coordinates. 
That is, the components of the correlation tensor are experimentally accessible by measuring the correlation function at the directions given by the basis vectors of local coordinate systems.
Obviously the assumed form of (\ref{et}) implies rotational invariance, because the correlation function is a scalar. Rotational invariance simply states that the value of $E(\vec n_1, \vec n_2, ..., \vec n_N)$ cannot depend on the local coordinate systems used by the $N$ observers. There is an important, although obvious,  implication of rotational invariance. Assume that  one knows the values of all $3^N$ components of the correlation tensor, $T_{i_1...i_N}$, which are obtainable by performing specific $3^N$  measurements of the correlation function, compare eq. (\ref{tensor}). Then, with the use of formula (\ref{et}) one can reproduce the value of the correlation functions for all other possible sets of local settings.

Using this rotationally invariant structure of the correlation function, 
we shall derive a necessary condition for the existence of local realistic 
description of the experimental correlation function (\ref{et}).

If the correlation function is described by a local realistic theory,
then the correlation function must be simulated by the following structure
\begin{eqnarray}
&&E_{LR}(\vec{n}_1,\vec{n}_2,\ldots,\vec{n}_N)=\nonumber\\
&&\int d\lambda \rho(\lambda)
I^{(1)}(\vec{n}_1,\lambda)I^{(2)}(\vec{n}_2,\lambda)\cdots
I^{(N)}(\vec{n}_N,\lambda),\label{LHVcofun}
\end{eqnarray}
where $\lambda$ is some local hidden variable, $\rho(\lambda)$ is a probabilistic distribution, and $I^{(j)}(\vec{n}_j,\lambda)$ is 
the predetermined ``hidden'' result of 
the measurement of the dichotomic observable $\vec n \cdot \vec \sigma$ with values $\pm 1$.

Let us parametrize the arbitrary unit vector in the plane defined by $\vec x_j^{(1)}$ and $\vec x_j^{(2)}$ in the following way
\begin{eqnarray}
\vec{n}_j(\alpha_j)=\cos \alpha_j \vec{x}_j^{(1)}
+\sin \alpha_j \vec{x}_j^{(2)}.\label{vector}
\end{eqnarray}
We shall show that the scalar product 
of any local realistic correlation function
\begin{eqnarray}
&&E_{LR}(\alpha_1,\alpha_2,\ldots,\alpha_N)=\nonumber\\
&&\int d\lambda \rho(\lambda)
I^{(1)}(\alpha_1,\lambda)I^{(2)}(\alpha_2,\lambda)\cdots
I^{(N)}(\alpha_N,\lambda),
\end{eqnarray}
with any rotationally invariant correlation function, that is
\begin{eqnarray}
&&E(\alpha_1,\alpha_2,\ldots,\alpha_N)=\nonumber\\
&&\hat{T} \cdot \vec{n}_1(\alpha_1)
\otimes
\vec{n}_2(\alpha_2)
\otimes\cdots
\otimes\vec{n}_N(\alpha_N),
\end{eqnarray}
is bounded by a specific number dependent on $\hat{T}$, namely:
\begin{eqnarray}
&& (E_{LR}, E) =\int_0^{2\pi}\!\!\!\!d\alpha_1
\int_0^{2\pi}\!\!\!\!d\alpha_2\cdots
\int_0^{2\pi}\!\!\!\!d\alpha_N\nonumber\\
&&E_{LR}(\alpha_1,\alpha_2,\ldots,\alpha_N)
E(\alpha_1,\alpha_2,\ldots,\alpha_N)\nonumber\\
&&\leq 4^N T_{max},\label{Bell-Zineq}
\end{eqnarray}
where $T_{max}$ is the maximal 
possible value of the correlation tensor component, i.e.
\begin{eqnarray}
T_{max}=\max_{\alpha_1,\alpha_2,\ldots,\alpha_N}
E(\alpha_1,\alpha_2,\ldots,\alpha_N).
\label{TE}
\end{eqnarray}

A necessary condition 
for the existence of a local realistic 
description $E_{LR}$ of the experimental correlation function 
\begin{equation}
E(\alpha_1,\alpha_2,\ldots,\alpha_N)=E(\vec n_1(\alpha_1),..., \vec n_N(\alpha_N)),
\end{equation} 
that is for $E_{LR}$ to be equal to $E$, 
is that one has $(E_{LR}, E) =(E, E)$.
If we have, e.g.,  $( E_{LR}, E)< (E, E)$,
then the experimental correlation 
function 
cannot be explainable by the local realistic theory. 

In what follows, we derive the upper bound (\ref{Bell-Zineq}).
Since the local realistic model is an average over $\lambda$, it is enough to find the bound of the following expression
\begin{eqnarray}
&&\int_0^{2\pi}\!\!\!\!d\alpha_1
\cdots
\int_0^{2\pi}\!\!\!\!d\alpha_N 
I^{(1)}(\alpha_1,\lambda)\cdots
I^{(N)}(\alpha_N,\lambda) \nonumber\\
&& \times \sum_{i_1,i_2,\ldots,i_N=1,2}T_{i_1i_2...i_N}
c^{i_1}_{1}c^{i_2}_{2}\cdots c^{i_N}_{N},\label{integral}
\end{eqnarray}
where
\begin{eqnarray}
(c^1_{j}, c^2_{j})=(\cos \alpha_j, \sin \alpha_j),
\end{eqnarray}
and
\begin{eqnarray}
T_{i_1i_2...i_N}=\hat{T} \cdot
(\vec{x}_1^{(i_1)}\otimes\vec{x}_2^{(i_2)}\otimes
\cdots\otimes\vec{x}_N^{(i_N)}),
\end{eqnarray}
compare (\ref{et}) and (\ref{tensor}). 

Let us analyze the structure of this integral (\ref{integral}).
Notice that (\ref{integral}) is a sum, with coefficients given by
$T_{i_1i_2...i_N}$, which is a product of the following integrals:
\begin{eqnarray}
\int_0^{2\pi}\!\!\!\!d\alpha_j I^{(j)}(\alpha_j, \lambda) \cos \alpha_j,
\end{eqnarray}
and
\begin{eqnarray}
\int_0^{2\pi}\!\!\!\!d\alpha_j I^{(j)}(\alpha_j, \lambda) \sin \alpha_j.
\end{eqnarray}
We deal here with integrals, or rather scalar products of
$I^{(j)}(\alpha_j, \lambda)$ with two orthogonal functions.
One has
\begin{eqnarray}
\int_0^{2\pi}\!\!\!\!d\alpha_j \cos \alpha_j\sin \alpha_j=0.
\end{eqnarray}
The normalized functions 
$\frac{1}{\sqrt{\pi}}\cos\alpha_j$ and $\frac{1}{\sqrt{\pi}}\sin\alpha_j$ form a basis of a real two-dimensional functional space, which we shall call $S^{(2)}$.
Note further that any function in $S^{(2)}$ is of the form 
\begin{eqnarray}
A\frac{1}{\sqrt{\pi}}\cos\alpha_j + B\frac{1}{\sqrt{\pi}}\sin\alpha_j,
\end{eqnarray}
where $A$ and $B$ are constants,
and that any normalized function in $S^{(2)}$ is given by
\begin{eqnarray}
&&\cos \psi \frac{1}{\sqrt{\pi}}\cos\alpha_j + \sin \psi \frac{1}{\sqrt{\pi}}\sin\alpha_j
\nonumber\\
&&=\frac{1}{\sqrt{\pi}}\cos(\alpha_j-\psi).
\end{eqnarray}
The norm 
$\Vert I^{(j){||}} \Vert$ of the projection 
of $I^{(j)}$ into the 
space $S^{(2)}$ is given by the maximal possible value of the scalar product $I^{(j)}$ with any normalized function belonging to $S^{(2)}$, that is
\begin{eqnarray}
\Vert I^{(j){||}} \Vert= \max_{\psi}
\int_0^{2\pi}\!\!\!\!d\alpha_j 
I^{(j)}(\alpha_j, \lambda)
\frac{1}{\sqrt{\pi}}\cos (\alpha_j-\psi).
\end{eqnarray}
Because $|I^{(j)}(\alpha_j, \lambda)|=1$,
one has $\Vert I^{(j){||}} \Vert\leq 4/\sqrt{\pi}$.

Since $\frac{1}{\sqrt{\pi}}\cos\alpha_j$ 
and $\frac{1}{\sqrt{\pi}}\sin\alpha_j$ are two orthogonal basis functions in $S^{(2)}$, one has
\begin{eqnarray}
\int_0^{2\pi}\!\!\!\!d\alpha_j 
I^{(j)}(\alpha_j, \lambda)
\frac{1}{\sqrt{\pi}}\cos \alpha_j=\cos \beta_j \Vert I^{(j){||}} \Vert,
\end{eqnarray}
and
\begin{eqnarray}
\int_0^{2\pi}\!\!\!\!d\alpha_j 
I^{(j)}(\alpha_j, \lambda)
\frac{1}{\sqrt{\pi}}\sin \alpha_j=\sin \beta_j \Vert I^{(j){||}} \Vert,
\end{eqnarray}
where $\beta_j$ is some angle.
Using this fact one can put the value of (\ref{integral}) into the following form
\begin{eqnarray}
&&(\sqrt{\pi})^{N}\prod_{j=1}^N \Vert I^{(j){||}} \Vert\nonumber\\
&& \times \sum_{i_1,i_2,\ldots,i_N=1,2}
T_{i_1i_2...i_N}
d^{i_1}_{1}d^{i_2}_{2}\cdots d^{i_N}_{N},
\label{EE}
\end{eqnarray}
where
\begin{eqnarray}
(d^{1}_j, d^{2}_j)=(\cos \beta_j, \sin \beta_j).
\label{dcos}
\end{eqnarray}

Let us look at the expression 
\begin{equation}
\sum_{i_1i_2...i_N=1,2}
T_{i_1i_2...i_N}
d^{i_1}_{1}d^{i_2}_{2}\cdots d^{i_N}_{N}.
\label{onT}
\end{equation}
Formula (\ref{dcos}) shows that we deal here with two dimensional unit vectors $\vec d_j = (d_j^1, d_j^2), j=1,2,...,N$, that is (\ref{onT}) is equal to $\hat T \cdot (\vec d_1 \otimes \vec d_2 \otimes .... \otimes \vec d_N)$, i.e. it is a component of the tensor $\hat T$ in the directions specified by the vectors $\vec d_j$. If one knows all the values of $T_{i_1i_2...i_N}$, one can always find the maximal possible value of such a component, and it is equal to $T_{max}$, of Eq.(\ref{TE}).

Therefore since $\Vert I^{(j){||}} \Vert\leq 4/\sqrt{\pi}$
the maximal value of (\ref{EE}) is $4^N T_{max}$, 
and finally one has
\begin{eqnarray}
(E_{LR}, E) \leq 4^N T_{max}.\label{fizu}
\end{eqnarray}

Please note that the relation (\ref{fizu}) is a generalized Bell inequality. All local realistic models must satisfy it (paradoxically, even those that are not leading to a rotationally invariant $E_{LR}$). Below we show that if one replaces $E_{LR}$ by $E$ one may have a violation of the inequality (\ref{fizu}). One has
\begin{eqnarray}
&&(E, E) = 
\int_0^{2\pi}\!\!\!\!d\alpha_1
\int_0^{2\pi}\!\!\!\!d\alpha_2\cdots
\int_0^{2\pi}\!\!\!\!d\alpha_N \nonumber\\
&& \times \left(\sum_{i_1,i_2,\ldots,i_N=1,2}T_{i_1i_2...i_N}
c^{i_1}_{1}c^{i_2}_{2}\cdots c^{i_N}_{N}\right)^2 \nonumber\\
&&=\pi^N \sum_{i_1,i_2,\ldots,i_N=1,2}T_{i_1i_2...i_N}^2.
\label{EEvalue}
\end{eqnarray}
Here, we have used the fact that $\int_{0}^{2\pi} d \alpha_j ~ c_j^{i_1} c_j^{i_1'} = \pi \delta_{i_1i_1'}$, because $c_j^1 = \cos \alpha_j$ and $c_j^2 = \sin{\alpha_j}$.
The structure of condition (\ref{fizu}) and the value (\ref{EEvalue}) suggests that the value of (\ref{EEvalue}) does not have to be smaller than (\ref{fizu}). That is there may be such correlation functions $E$, which have the property that for any $E_{LR}$ one has $(E_{LR}, E) < (E,E)$, which implies impossibility of modeling $E$ by any local realistic correlation function $E_{LR}$.

We present here an example of violation of the inequality (\ref{fizu}). Imagine $N$ observers who can choose between two orthogonal directions of spin measurement, $\vec x_j^{(1)}$ and $\vec x_j^{(2)}$ for the $j$th one. Let us assume that the source of $N$ entangled spin-carrying particles emits them in a state, which can be described as $V |\psi_{GHZ} \rangle \langle
\psi_{GHZ}|+(1-V) \rho_{noise}$, where $|\psi_{GHZ}
\rangle = 1/\sqrt{2} (|+\rangle_1\cdots|+\rangle_N +
|-\rangle_1 \cdots|-\rangle_N)$ is the 
Greenberger, Horne, and Zeilinger (GHZ) state \cite{bib:GHZ} and
$\rho_{noise} = \frac{1}{2^N} \openone$ is the random noise admixture. The value of $V$ can be interpreted as the reduction factor of the interferometric contrast observed in the multi-particle correlation experiment. The states $| \pm \rangle_j$ are the eigenstates of the $\sigma_z^j$ observable. One can easily show that if the observers limit their settings to $\vec x_j^{(1)} = \hat x_j$ and $\vec x_j^{(2)} = \hat y_j$ there are $2^{N-1}$ components of $\hat T$ of the value $\pm V$. These are $T_{11...1}$ and all components that except from indices 1 have an even number of indices 2. Other x-y components vanish.

It is easy to see that $T_{max}=V$ and 
$\sum_{i_1,i_2,\ldots,i_N=1,2}T_{i_1i_2...i_N}^2=V^2 2^{N-1}$.
Then, we have $(E_{LR}, E) \leq 4^N V$ and $(E,E) = \pi^N V^2 2^{N-1}$.
If one has  more than three spins the rotational invariance puts an additional, non trivial, constraint on the local realistic models. For $N \geq 4$, and $V$ given by
\begin{equation}
2(2/\pi)^N<V \leq \frac{1}{\sqrt{2^{N-1}}}
\end{equation}
despite the fact that there exists a local realistic model for the actually measured values of the correlation function, the rotational invariance principle disqualifies this model. As it was shown in \cite{bib:Zukowski2} if the correlation tensor satisfies the following condition
\begin{equation}
\sum_{i_1,i_2,...,i_N=1,2} T_{i_1i_2...i_N}^2 \leq 1
\label{ZB}
\end{equation}
then there always exists an {\em explicit} local realistic model for the set of correlation function values $E(\vec x_{1}^{(i_1)},\vec x_{2}^{(i_2)},...,\vec x_{N}^{(i_N)})$, $i_1,i_2,...,i_N=1,2$. For our example the condition (\ref{ZB}) is met whenever $V \leq \frac{1}{\sqrt{2^{N-1}}}$. Nevertheless the rotational invariance principle excludes local realistic models for $V > 2(\frac{2}{\pi})^N$. 
Thus the situation is such: for $V \leq \frac{1}{\sqrt{2^{N-1}}}$ for all two settings per observer
experiments  one can construct a local realistic
model for the values of the correlation function for the settings chosen in the experiment. But these models must be consistent with each other,
if we want to extend their validity beyond the $2^N$ settings to which each
of them pertains.
Our result clearly indicates that this is impossible for $V > 2(\frac{2}{\pi})^N$.
That is, models built to reconstruct the $2^N$ data points, when compared
with each other, must be inconsistent - therefore they are invalidated.
The models must contradict each other.
In other words the explicit models, given in \cite{bib:Zukowski2}
work only for the specific set of settings in the given experiment, but
cannot be extended to all settings. We utilize rotational invariance to
show this.

Please note that all information needed to get this conclusion can be obtained in a two-orthogonal-settings-per-observer experiments, that is with the information needed in the case of ``standard'' two settings Bell inequalities \cite{bib:Mermin, bib:Zukowski2, bib:Werner3}. To get both the value of (\ref{EEvalue}) and of $T_{max}$ it is enough to measure all values of $T_{i_1i_2...i_N}$, $i_1, i_2, ..., i_N=1,2$. 

In summary we have shown that if except for the assumptions of locality and realism, one also requires that the correlation functions in a Bell experiment have an explicitly rotationally invariant form, this leads, in some important cases to a stronger version of Bell's theorem. The interesting feature is that Bell's theorem rules out realistic interpretation of some quantum mechanical predictions, and therefore of quantum mechanics in general, provided one assumes locality. Locality is a consequence of the general symmetries of the Poincar{\'e} group of the Special Relativity Theory. However it is a direct consequence of the Lorentz transformations (boosts), as they define the light-cone. As our discussion shows a subgroup of the Poincar{\'e} group, rotations of the Cartesian coordinates, introduces an additional constraint on the local realistic models. 

The interesting feature of this additional constraint of rotational invariance is that it is introduced on the level of correlation functions, i.e. formally speaking after averaging over hidden variables ($\lambda$'s). In contradistinction the locality condition is here introduced for every value of $\lambda$, i.e. by assuming that the local result are determined only by $\lambda$ and the local setting (what is mathematically expressed by the postulate of existence of functions $I^{(j)}(\vec n_j, \lambda)$, compare (\ref{LHVcofun})). It is well known that if one assumes locality only for the averaged hidden variable theories, i.e. on the level of correlation functions, one gets the so called ``no-signaling condition''. Such a condition does not rule out a realistic interpretation of quantum mechanics (since one has e.g. Bohm's model).

It would be very interesting to consider situations in which other symmetries of physical laws constrain additionally local realistic theories.

Finally we would like to mention a related preprint that recently appeared \cite{bib:U}, which utilizes the conservation laws as additional constraints. However, the approach of \cite{bib:U} (for two qubits) requires perfect spin anticorrelations, i.e. something that is experimentally impossible to realize.

We would like to acknowledge an important remark by prof. N. D. Mermin concerning the first version of the paper.
M.Z. is supported by the Professorial Subsidy of FNP. W.L. and M.W. are supported by FNP and UG grant BW-5400-5-0260-4. The work is part of the MNiI project 1 P03B 04927.

\end{document}